\documentclass[12pt]{article}
\pdfoutput=1
\usepackage{amssymb,amsmath,xcolor,empheq}
\numberwithin{equation}{section}
\usepackage{epsfig}
\usepackage{epstopdf}
\usepackage{latexsym}
\usepackage{graphicx}
\usepackage{booktabs}
\usepackage{bbm}
\usepackage{enumitem}
\usepackage[numbers,compress]{natbib}
\usepackage{bm}
\usepackage[T1]{fontenc}
\usepackage{amsthm}

\usepackage{natbib}
\usepackage{fancybox}

\usepackage[margin=20pt,small]{caption}
\usepackage{subcaption}

\usepackage[toc]{appendix}

\usepackage{color}
\usepackage{datetime}
\usepackage[
      colorlinks=false,
      linkcolor=darkblue,  
      urlcolor=blue,    
      filecolor=blue,     
      citecolor=red,
linktocpage=true,
      pdfstartview=FitV,
      bookmarksopen=true,
	  hidelinks
      ]{hyperref}

\ifpdf
\fi

\newcommand*{\boxedcolor}{black}
\makeatletter
\renewcommand{\boxed}[1]{\textcolor{\boxedcolor}{%
  \fbox{\normalcolor\m@th$\displaystyle#1$}}}
\makeatother

\usepackage{calligra}
\DeclareMathAlphabet{\mathcalligra}{T1}{calligra}{m}{n}

\definecolor{cardinal}{rgb}{0.6,0,0}
\definecolor{darkgreen}{rgb}{0,0.5,0}
\definecolor{golden}{rgb}{0.92, 0.7, 0}
\definecolor{midnight}{rgb}{0, 0, 0.5}
\definecolor{darkblue}{rgb}{0.2, 0, 0.8}

\def\cA{{\cal A}}

\def\cD{{\cal D}}

\def\cI{{\cal I}}

\def\cN{{\cal N}}

\def\cQ{{\cal Q}}
\def\cR{{\cal R}}

\def\SU{{\rm SU}}
\def\U{{\rm U}}
\def\SO{{\rm SO}}
\def\rmi{{\rm i}}

\def\SL{{\rm SL}}

\def\SO{{\rm SO}}

\def\SU{{\rm SU}}

\def\fg{\frak{g}}

\def\cals#1{\mathcal{#1}}

\def\cA{{\cals A}}

\newcommand{\be}{\begin{equation}} \newcommand{\ee}{\end{equation}}
\newcommand{\bea}{\begin{equation} \begin{aligned}} \newcommand{\eea}{\end{aligned} \end{equation}}
\newcommand{\bmu}{\begin{multline}} \newcommand{\emu}{\end{multline}}

\newcommand\equ[1] {\begin{equation}#1\end{equation}}

\newcommand\eqs[1] {\begin{align}#1\end{align}}
\newcommand\eqss[1] {\begin{align}\begin{split}#1\end{split}\end{align}} 

\renewcommand\( {\left(}
\renewcommand\) {\right)}

\usepackage[cmtip,all]{xy}
\newcommand{\longsquiggly}{\xymatrix{{}\ar@{~>}[r]&{}}}

\begin{document}  

\begin{titlepage}

\begin{center} 

\vspace*{1.5cm}

{\huge  Universal Spinning Black Holes \\ \vspace{5mm}
and Theories of Class $\mathcal R$}

\vspace{1.5cm}

{\large  Nikolay Bobev${}^{(1)}$ and P. Marcos Crichigno${}^{(2)}$   \\ }
\bigskip
\bigskip
${}^{(1)}$
Instituut voor Theoretische Fysica, KU Leuven \\
Celestijnenlaan 200D, B-3001 Leuven, Belgium
\vskip 5mm
${}^{(2)}$ Institute for Theoretical Physics, University of Amsterdam,\\ 
Science Park 904, Postbus 94485, 1090 GL, Amsterdam, The Netherlands
\vskip 5mm
\texttt{nikolay.bobev@kuleuven.be,~p.m.crichigno@uva.nl} \\
\end{center}

\vspace{0.5cm}

\noindent 

\abstract{We study a supersymmetric, rotating, electrically charged black hole in AdS$_{4}$ which is a solution of four-dimensional minimal gauged supergravity. Using holography we show that the free energy on $S^3$ and the superconformal index of the dual three-dimensional $\cN=2$ SCFT, in the planar limit, are related in a simple universal way. This result applies to large classes of SCFTs constructed from branes in string and M-theory which we discuss in some detail. For theories of class $\cR$, which arise from $N$ M5-branes wrapped on hyperbolic three-manifolds, we show that the superconformal index agrees with the black hole entropy in the large $N$ limit. }

\end{titlepage}


\setcounter{tocdepth}{2}
\tableofcontents


\section{Introduction}

The microscopic counting of black hole entropy is one of the greatest accomplishments of string theory. This was first achieved for five-dimensional black holes in asymptotically flat spacetime in the seminal work  \cite{Strominger:1996sh}. For black holes in asymptotically locally AdS spacetimes this has only been recently achieved, starting with certain magnetic black holes in 4d in \cite{Benini:2015eyy},  extended in various ways in \cite{Benini:2016rke, Azzurli:2017kxo, Cabo-Bizet:2017jsl, Benini:2017oxt,Hosseini:2017fjo}, for magnetic black holes in 6d in \cite{Crichigno:2018adf,Hosseini:2018uzp,Fluder:2019szh,Hosseini:2018usu,Suh:2018tul,Suh:2018szn}, and for spinning black holes in 5d in \cite{Hosseini:2017mds,Cabo-Bizet:2018ehj,Choi:2018hmj,Benini:2018ywd,Honda:2019cio,ArabiArdehali:2019tdm}. These calculations  rely on holography, exploiting the properties of the asymptotic SCFT dual and the powerful supersymmetric localization techniques developed in the past decade.  See \cite{Zaffaroni:2019dhb} for a comprehensive review of these developments and a more complete list of references. 

In this paper we continue this program and consider a well-known black hole solution - the four-dimensional Kerr-Newman black hole in AdS. The general solution was first found in the late-1960's  by Carter \cite{Carter:1968ks}. Here we study a BPS limit of this background such that it can be viewed as a supersymmetric solution of 4d minimal $\cN=2$ gauged supergravity preserving two real supercharges  \cite{Kostelecky:1995ei,Caldarelli:1998hg}. This supersymmetric black hole has a nontrivial entropy and carries both angular momentum and electric charge. Its mass is related to these charges by the usual BPS relation. 

To account for the microscopic origin of the black hole entropy in string or M-theory one must first embed the solution into 10d or 11d supergravity. There are infinitely many ways of doing so, specified by a choice of internal manifold and the fluxes on it. This is because 4d $\cN=2$ minimal gauged supergravity arises as a consistent truncation of 10d or 11d supergravity on certain internal manifolds, $M_6$ or $M_7$, respectively. The precise microscopic interpretation of the black hole in string or M-theory depends on the D- or M-brane realization of the supergravity solution, which in turn also determines the 3d $\cN=2$ SCFT living at the asymptotic boundary. The general expectation is that the entropy of the black hole is captured by the degeneracy of  states in the 3d SCFT which preserve the same amount of supersymmetry as the black hole and carry the same charges. This is encoded in the superconformal index $\cI_{S^2}$,  or $S^2\times S^1$ partition function, of the theory \cite{Bhattacharya:2008zy,Imamura:2011su}. By the state-operator correspondence the superconformal index can also be seen as counting local operators of the conformal theory in flat space. Since minimal\footnote{Here ``minimal'' has the same meaning as in \cite{Bobev:2017uzs}, i.e., the gauged supergravity theory with a fixed number of supercharges that contains only the  gravity multiplet.}  4d gauged supergravity contains only the gravity multiplet, which is dual to the energy-momentum multiplet in the 3d SCFT,  the microstate degeneracy must  be captured by the degeneracy of operators with a given superconformal R-charge and angular momentum, irrespective of their charge under other potential flavor symmetries in the field theory.

We   refer to this supergravity solution as a ``universal spinning black hole,'' following an analogous discussion for static magnetic black holes  in AdS$_4$ \cite{Azzurli:2017kxo} and, more generally, for  static black $p$-brane solutions in various dimensions \cite{Bobev:2017uzs,Benini:2015bwz}. As we discuss in detail, this universality amounts to an interesting consequence for the behavior of the superconformal index  of {\it any} 3d $\cN=2$ SCFT with a weakly coupled gravity dual in the large $N$ limit. Namely, in a regime in which the universal spinning black hole solution is the dominant contribution to the index, it follows that to leading order in $N$,
\equ{\label{IS3intro}
\log \cI_{S^2}(\varphi, \omega)\approx \rmi\frac{F_{S^{3}}}{\pi } \frac{\varphi^{2}}{\omega}\,,
}
where $\omega$ and $\varphi$  are fugacities for rotations of the  $S^2$  and for the  superconformal $\U(1)_R$ R-symmetry, respectively. The round-sphere free energy, $F_{S^{3}}$,  appearing here is evaluated at the  {\it superconformal} values of the R-charges, which can be determined by F-extremization \cite{Jafferis:2010un}. We note that this expression is reminiscent of the Cardy formula for 2d CFTs, where $F_{S^3}$ here plays the role of the 2d central charge. If the theory has global flavor symmetries one can refine the index by including  fugacities and magnetic fluxes through the $S^2$ for these symmetries \cite{Kapustin:2011jm}.  As mentioned above, however, our focus here  is on the universal case for which all flavor parameters are turned off.\footnote{The flavor symmetry in the field theory is realized by gauge fields sitting outside the 4d gravity multiplet which are not included in the universal field content of the minimal supergravity theory. The generalized index with flavor fugacities and magnetic fluxes should account for the entropy of black holes with additional electric and magnetic charges. } A discrete  refinement of the index, which {\it is} relevant in our discussion, is related to a choice of spin structure on $S^2\times S^1$ \cite{Closset:2018ghr}. We show that to account for the black hole degeneracy one must choose {\it anti-periodic} boundary conditions for fermions around the $S^1$, which implies the relation among fugacities $\omega-2\varphi \pm 2\pi \rmi=0$. This is analogous to a recent analysis for supersymmetric rotating black holes in AdS$_5$ \cite{Cabo-Bizet:2018ehj}.

On general grounds, the superconformal index  takes the form
\equ{\label{eq:Iintro}
\cI_{S^2}(\varphi, \omega)=\sum_{Q,J}\Omega(Q,J)\,e^{\varphi \, Q}\, e^{\omega J} \,,
}
where the sum is over states in the Hilbert space of the theory quantized on $S^2$ preserving two supercharges,  with R-charge $Q$ and angular momentum $J$ (see Appendix~\ref{sec:SCindex}). The coefficients  $\Omega(Q,J)$ count the degeneracy of such states, which can be extracted from the index by the inverse transform. Schematically,
\equ{\label{eq:Omintro}
\Omega(Q,J)=\int_{\mathcal C}\frac{d\varphi}{2\pi \rmi }\frac{d\omega}{2\pi \rmi }\, \cI_{S^2}(\varphi, \omega)\, e^{-\omega J}\, e^{-\varphi \, Q}\approx e^{\log \cI_{S^2}(\varphi,\, \omega)-\varphi \, Q-\omega J}\Big|_{\text{s.p.}}\,,
}
where $\mathcal C$ is a suitable contour and  $(\cdots)|_{\text{s.p.}}$ stands for evaluating the function in the saddle-point values, which dominate the integral in the large charge limit where the supergravity approximation is valid.\footnote{To be more precise, in this procedure one must take into account the constraint among fugacities which can be done by including an integration over a Lagrange multiplier, $\int \frac{d\lambda}{2\pi i} e^{\lambda (\omega-2\varphi \pm 2\pi \rmi)}$, as discussed in Section~\ref{sec:Supersymmetry and extremality}.  }  As discussed in more detail below, this procedure leads to a complex  $\Omega(Q,J)$. However, regularity of the black hole imposes an additional constraint $J=J(Q)$ in which case $\Omega(Q,J)$ becomes real and to leading order in $N$ reproduces the macroscopic entropy:
\equ{\label{eq:SBHintro}
S_{\text{BH}}= \log \Omega(Q,J(Q))\,.
}

Thus, counting  the number of microstates of the universal spinning black hole, irrespective of its particular uplift to string or M-theory,  amounts to establishing \eqref{IS3intro} for generic 3d $\cN=2$ SCFTs with a weakly coupled gravity dual in the large $N$ limit. For 3d SCFTs with a gauge theory description in the UV, both sides of \eqref{IS3intro} can be computed via supersymmetric localization. Thus, a direct check of this relation is in principle possible for these theories. However, the superconformal index is a rather complicated object and it may not be straightforward to evaluate it in the large $N$ limit and establish this general behavior. We note that the large $N$ behavior of the superconformal index of the ABJM theory was studied in the presence of general flavor fugacities in \cite{Choi:2019zpz}. When the results of \cite{Choi:2019zpz} are specialized to the universal setup we study here one recovers the relation in \eqref{IS3intro}.

To gain nontrivial evidence for the validity of \eqref{IS3intro} in a large class of SCFTs we study 3d theories of class $\mathcal{R}$ obtained by twisted compactification of the 6d $(2,0)$ $A_{N-1}$ theory on a hyperbolic three manifold $\Sigma_3$. Although these theories generically do not have Lagrangian descriptions in 3d one can exploit the 6d origin of the theory by using the 3d-3d correspondence  relating the superconformal index of the theory to certain topological invariants of $\Sigma_3$. See \cite{Dimofte:2011ju,Cecotti:2011iy,Dimofte:2011py,Dimofte:2010tz,Terashima:2011qi,Dimofte:2011jd} and \cite{Dimofte:2014ija} for a review. Using large $N$ results for these invariants derived in \cite{Gang:2014ema} we are able to establish the relation in \eqref{IS3intro} for this class of theories. This leads, in particular, to a microscopic counting of the entropy of the corresponding supersymmetric universal spinning black hole arising from M5-branes wrapped on $\Sigma_3$. 

The paper is organized as follows. In Section~\ref{sec:A universal spinning black hole in  AdS4} we review the Kerr-Newman-AdS black hole paying particular attention to its thermodynamics and its supersymmetric and extremal limits. In Section~\ref{sec:Uplifts and universality} we discuss several distinct explicit uplifts of this solution to string and M-theory. In Section~\ref{sec:Microscopic entropy} we discuss the microscopic counting of the black hole entropy for the uplift to M-theory arising from M5-branes wrapping a hyperbolic three manifold. We conclude in Section~\ref{sec:Outlook} with a discussion of some open problems. The two Appendices contain some details on the superconformal index as well as the Killing spinors preserved by the black hole solution.

\section{A universal spinning black hole in  AdS$_{4}$}
\label{sec:A universal spinning black hole in  AdS4}

Consider four-dimensional minimal $\cN=2$ gauged supergravity, with bosonic field content the graviton and a $\U(1)$ graviphoton. The bosonic action is given by 
\equ{\label{Sugra_action4d}
I=\frac{1}{16\pi G_{(4)}}\int d^{4}x \sqrt{-g}\(R+6-\frac14 F^{2}\)\,,
}
with $G_{(4)}$ the four-dimensional Newton constant. The equations of motion are
\eqss{\label{EOM4d}
R_{\mu\nu}+3 g_{\mu\nu}-\frac{1}{2}\, (F_{\mu\sigma}F_{\nu}\,^{\sigma}-\frac{1}{4} g_{\mu\nu}F_{\rho\sigma}F^{\rho\sigma})=0\,,\\
\partial_{\mu}F^{\mu\nu}=0\,.
}
We have fixed the value of the cosmological constant so that the AdS$_{4}$ solution  has radius $L_{\text{AdS}_{4}}=1$.

The equations of motion \eqref{EOM4d} admit a spinning, electrically charged black hole solution \cite {Carter:1968ks} (see also \cite{Kostelecky:1995ei,Caldarelli:1998hg}). In Lorentzian, mostly plus signature it is given by\footnote{Here we follow the conventions of \cite{Cvetic:2005zi} and set, in the notation there,  $\delta_{1}=\delta_{2}\equiv \delta$ to truncate to four-dimensional minimal $\cN=2$ gauged supergravity.}
\eqss{\label{BHsol}
ds^{2}_{4}=\,&-\frac{\Delta_{r}}{W}\(dt-\frac{a \sin^{2}\theta}{\Xi} d\phi\)^{2}+W\(\frac{dr^{2}}{\Delta_{r}}+\frac{d\theta^{2}}{\Delta_{\theta}}\)\,\\
\,&\hspace{40mm}+\frac{\Delta_{\theta}\sin^{2}\theta}{W}\(a dt -\frac{(\tilde r^{2}+a^{2})}{\Xi}d\phi\)^{2} \,, \\
A=\,&\frac{ 2 m \tilde r \sinh 2\delta }{W}\(dt-\frac{a \sin^{2}\theta}{\Xi} d\phi\) +\alpha dt\,,
}
where we have defined 
\eqss{\label{eq:metfuncBH}
\tilde r=&\,r+2m\sinh^{2}\delta\,, \qquad \Delta_{r}=\,r^{2}+a^{2}-2m r+\tilde r^{2}(\tilde r^{2}+a^{2})\,,\\
\Delta_{\theta}=\,&\,1-a^{2}\cos^{2}\theta\,,\qquad W=\tilde r^{2}+a^{2}\cos^{2}\theta\,,\qquad \Xi=1-a^{2}\,.
}
The solution is specified by the three integration constants, $(a,\delta,m)$ and $\alpha$. The parameter $\alpha$ does not affect the metric, being a pure gauge transformation of the gauge field, but is nonetheless important as it fixes the spin structure of the asymptotic boundary, as we discuss below. The solution describes an  AdS black hole with an outer and an inner horizon, provided $m$ is larger than a critical value and $a^{2}<1$ (see, e.g., \cite{Caldarelli:1999xj}). Without loss of generality we can assume $a\geq0,\delta\geq 0,m\geq 0$.\footnote{The sign of $a$ can be absorbed by the coordinate redefinition, $\phi\to -\phi$,  the sign of $\delta$  can be absorbed by sending $A\to -A$ in the action, and the sign of $m$ can be absorbed by redefining the radial coordinate, $r\to -r$. } 
 The physical quantities characterizing the black hole are its energy $E$, electric charge $Q$, and angular momentum $J$, given by\footnote{Note we define $Q$ with a factor of 4 difference compared to \cite{Cvetic:2005zi}.}
\equ{\label{EQJ}
E=\frac{m}{G_{(4)}\Xi^{2}}\cosh 2\delta\,,\qquad Q=\frac{m}{G_{(4)}\Xi}\sinh 2\delta\,,\qquad J=\frac{ma}{G_{(4)}\Xi^{2}}\cosh 2\delta\,.
}
Since supergravity is reliable at weak gravitational coupling, for generic values of the parameters $a,\delta,m$, the charges \eqref{EQJ} are large in the classical supergravity regime. The Bekenstein-Hawking entropy is given by the area of the outer horizon and reads
\equ{
S=\frac{\text{Area}}{4G_{(4)}}=\frac{\pi(\tilde r^{2}+a^{2})}{G_{(4)}\Xi}\Big|_{r=r_{+}}\,,
}
where $r_+$ denotes the location of the outer horizon and is given by the largest real positive root of the quartic polynomial equation $\Delta_r=0$. 

The metric can be analytically continued to Euclidean signature by introducing the Euclidean time $\tau= \rmi t $ and continuing $a$ to purely imaginary values. Demanding  regularity of the Euclidean metric at the horizon implies the identifications \cite{Caldarelli:1999xj} 
\equ{\label{tauphi}
(\tau,\phi)\sim (\tau+\beta,\phi- \rmi \beta \Omega_H)\,,
}
where $\beta\equiv T^{-1}$ is the inverse temperature of the black hole and $\Omega_H$ is the angular velocity of the horizon, given by
\equ{\label{T}
T  =\frac{1}{4\pi (\tilde r^{2}+a^{2})}\frac{d\Delta_{r}}{dr}\Big|_{r=r_+}\,, \qquad \Omega_H= \frac{a\, \Xi}{\tilde r^2+a^2}\Big|_{r=r_{+}}\,.
}
To derive \eqref{tauphi} and \eqref{T} it is useful to analytically continue the metric in \eqref{BHsol} after first rewriting it as
\eqss{
ds^{2}_{4}=\,&-\frac{W\Delta_{r}\Delta_{\theta}}{\Sigma} dt^{2}+W\(\frac{dr^{2}}{\Delta_{r}}+\frac{d\theta^{2}}{\Delta_{\theta}}\)\,\\
\,&\hspace{40mm}+\frac{\Sigma\sin^2\theta}{W\Xi^2}\(d\phi -\frac{a\Xi}{\Sigma}(\Delta_{\theta}(\tilde{r}^2+a^2)-\Delta_r)dt\)^{2} \,,
}
where 
\begin{equation}
\Sigma = (\tilde{r}^2+a^2)^2\Delta_{\theta}-\Delta_r a^2\sin^2\theta\,.
\end{equation}
The thermodynamics of this black hole was discussed in some detail in \cite{Caldarelli:1999xj} and more recently in \cite{Chow:2013gba,Choi:2018fdc,Cassani:2019mms}.  One can define chemical potentials for the angular momentum and the gauge field, which are given by
\eqs{\label{eq:OmPhidef}
\Omega=\frac{a (1+\tilde r^{2})}{\tilde r^{2}+a^{2}}\Big|_{r=r_{+}}\,,\qquad 
\Phi=\frac{m  \tilde r \sinh 2\delta }{\tilde r^{2}+a^{2}}\Big|_{r=r_{+}}\,,
}
respectively. With this at hand it can be shown that the black hole obeys the first law of thermodynamics,
\equ{
dE=T \, dS +\Omega \, dJ+\Phi \, dQ\,.
}
%

\subsection{Asymptotics}

Let us first note that setting $m=\delta=a=0$ the gauge field is pure gauge and after the change of coordinates $r=\sinh\rho$ the  Euclidean metric reduces to 
\begin{equation}
ds^2_4 = \cosh^2\rho \; d\tau^2 +d\rho^2 +\sinh^2\rho \; (d\theta^2+\sin^2\theta \;d\phi^2)\,.
\end{equation}
This is the metric of the unit radius global AdS$_4$ with an $\Bbb R\times S^2$ asymptotic boundary, which one can compactify to  $S^1\times S^2$. Alternatively, one can make a conformal transformation of the metric to have a Euclidean AdS$_4$ solution with an $S^3$ boundary. The regularized on-shell action of this Euclidean solution, $I_{S^{3}}$,  is identified with the $S^3$ free energy, $F_{S^3}$, of the holographically dual CFT which takes the simple form \cite{Emparan:1999pm}
\equ{\label{FS3GN}
F_{S^{3}}= \frac{\pi }{2 G_{(4)}}\,.
}

For arbitrary value of the parameters $(a,\delta,m)$ the asymptotic boundary is $S^1\times S^2$ locally but not globally, as a consequence of imposing regularity of the Euclidean black hole solution close to the horizon.  To see this it is convenient to make a change of coordinates from $(r,\theta,\phi)$ to   $(\hat r, \hat \theta\,,\hat \phi)$ by writing, as in \cite{Hawking:1998kw}, 
\equ{
\hat r \cos \hat \theta = r \cos \theta\,,\qquad \hat r^{2}=\frac{1}{\Xi}\left[r^{2}\Delta_{\theta}+a^{2}\sin^{2}\theta\right]\,, \qquad  \phi = \hat \phi+ \rmi a \tau\,.
}
Then, in the limit $\hat r\to \infty$ the metric and gauge field asymptote to
 \eqss{\label{bdrymetriclocal}
 ds^2_4\,&\approx  \frac{d\hat r^2}{\hat r^2} +\hat r^2ds^2_{\text{bdry}}\,, \qquad ds^2_{\text{bdry}}= d\tau^2+d\hat \theta^2+\sin^2\hat \theta\;d\hat \phi ^2\,,\\
A\,&\approx -\rmi \alpha \, d\tau\,,
 }
where the boundary metric $ds^2_{\text{bdry}}$ is  the canonical metric on locally $S^1\times S^2$. Note that in terms of the new angular variable $\hat \phi$ the identification \eqref{tauphi} becomes
\equ{\label{idphihat}
(\tau,\hat \phi)\sim (\tau+\beta,\hat \phi- \rmi \beta \Omega)\,.
}
Alternatively, one can define $\tilde  \phi= \hat \phi+ \rmi \Omega \tau$ so that
 \equ{\label{bdrymetricglobal}
ds^2_{\text{bdry}}=d\tau^2+d\hat \theta^2+\sin^2\hat \theta\;( d\tilde \phi- \rmi \Omega d\tau)^2\,,
 }
 describing a fibration of $S^2$ over $S^1$. In these coordinates going around the Euclidean time circle is described by the identification $(\tau,\tilde \phi)\sim (\tau+\beta,\tilde \phi)$ while going around the angular coordinate  $(\tau,\tilde \phi)\sim(\tau, \tilde\phi+2\pi)$.

\subsection{Supersymmetry and extremality}
\label{sec:Supersymmetry and extremality}

The black hole solution reviewed above admits two important limits; the supersymmetric limit and the extremal limit  \cite{Kostelecky:1995ei,Caldarelli:1998hg}. The BPS limit is defined by first imposing supersymmetry  and then imposing extremality.  The supersymmetric limit is achieved by requiring 
\equ{\label{eq:susylimit}
e^{4\delta}=1+\frac{2}{a }\qquad \Rightarrow \qquad  E= J+ Q\,,
} 
leaving two independent parameters and thus two independent physical quantities, which we take to be $Q$ and $J$. Note that since in Euclidean signature $a$ is taken to purely imaginary, this requires a complex $\delta$.  One can show that supersymmetry implies a constraint among the chemical potentials $\Omega, \Phi$  \cite{Cassani:2019mms}:
\equ{\label{constOmegaPhi}
\beta (1+\Omega -2 \Phi)=\pm 2 \pi \rmi \,
}
or, after defining 
\begin{equation}\label{eq:omphidefsugra}
\omega\equiv \beta (\Omega-1)\,, \qquad \varphi\equiv\beta (\Phi-1)\,,
\end{equation}
the constraint is 
\equ{\label{constOP}
\omega-2\varphi =\pm 2\pi  \rmi \,.
}
The choice of sign arises since, after imposing the supersymmetric constraint in \eqref{eq:susylimit}, the metric function $\Delta_{r}$ in \eqref{eq:metfuncBH} does not generically have real zeros but rather two sets of complex conjugate zeroes. One can then choose either one of these roots to play the role of $r_+$ in \eqref{eq:OmPhidef}. The upper sign in \eqref{constOP} corresponds to choosing one of the complex roots and the lower sign to choosing its complex conjugate. From now on we choose the lower sign in \eqref{constOP}. 

As we show in Appendix~\ref{sec:Boundary Killing spinors} the constraint \eqref{constOP} determines the spin structure on the $S^1\times S^2$ asymptotic boundary, forcing the spinors generating the preserved supersymmetry to be {\it anti-periodic} when going around the Euclidean time circle, rather then the more standard periodic condition. This is analogous to the case of supersymmetric spinning black holes in AdS$_5$  discussed in \cite{Cabo-Bizet:2018ehj}.

In the supersymmetric limit,
\equ{\label{sugrawphi}
\omega=\frac{2\pi \rmi (a-1)}{1+a+2 \rmi \tilde r}\Big|_{r=r_{+}}\,,\qquad \varphi=\frac{2\pi \rmi (a+\rmi \tilde r)}{1+a+2 \rmi \tilde r }\Big|_{r=r_{+}}\,.
}

\

Another quantity of physical interest is the Euclidean on-shell action, $I_{E}$, for the supersymmetric Euclidean saddle-points obtained in the limit \eqref{eq:susylimit}. After appropriate holographic renormalization $I_E$ can be written as a function of the chemical potentials $\varphi$, $\omega$ as\footnote{This follows from the results in \cite{Cassani:2019mms}, setting $\delta_{1}=\delta_{2}$ to truncate to the minimal theory and reinstating $G_{(4)}$, which was set to 1 there. } 
\equ{\label{eq:IEsusy}
I_{E}=-\frac{\rmi}{2G_{(4)}} \frac{\varphi^{2}}{\omega}\,.
}
Note that in general this is a complex function. It was shown in \cite{Cassani:2019mms} that for the Euclidean supersymmetric solutions this on-shell action obeys the following quantum statistical relation 
\begin{equation}\label{eq:QSR}
S=-I_E-\omega J-\varphi \, Q\,.
\end{equation}

An  interesting property of the function \eqref{eq:IEsusy} is that it serves as an ``entropy function'' for the BPS black hole \cite{Choi:2018fdc,Cassani:2019mms}. More precisely, one should consider $\varphi,\omega, Q,J$ as independent parameters and define the auxiliary function
\equ{
 S_\lambda(Q,J;\varphi, \omega)\equiv -I_E -\omega J-\varphi \, Q+\lambda(\omega-2\varphi + 2\pi  \rmi)\,.
}
Then, extremizing $S_\lambda$ with respect to $\omega, \varphi$ and $\lambda$ gives the equations
\equ{\label{eqsLeg}
\frac{\partial I_{E} }{\partial \omega}=-J+\lambda\,,\qquad \frac{\partial I_{E} }{\partial \varphi}=-Q-2\lambda\,,\qquad \omega-2\varphi+ 2\pi \rmi=0\,.
}
One can easily check that the values \eqref{sugrawphi} are solutions to these equations and thus are extrema of  $S_\lambda$. Furthermore, the extremal value of $S_\lambda$ coincides with the supersymmetric entropy $S$, which is automatic by virtue of \eqref{eq:QSR}.

\

According to the AdS/CFT dictionary, $I_E$ should be compared to the supersymmetric partition function, $Z_{S^{2}_{\omega}\times S^{1}}$,  of the 3d $\cN=2$ SCFT living at the $S^{2}_{\omega} \times S^{1}$ asymptotic boundary:
\begin{equation}\label{eq:IZI}
\cI_{S^2}(\omega) = Z_{S^{2}_{\omega}\times S^{1}}\approx e^{-I_{E}}\,.
\end{equation}
This relation is valid to leading order in the large $N$ limit and we have included the label $\omega$ to indicate that the $S^2$ is fibered over the $S^1$ and the field theory partition function has to be evaluated with fugacities $\omega$ and $\varphi$ obeying the relation  \eqref{constOP}. This partition function is also referred to as the superconformal index.

Combining the relations \eqref{FS3GN} and \eqref{eq:IEsusy} with \eqref{eq:IZI} the supergravity calculations above lead to the following prediction for the large $N$ limit of the superconformal index of all 3d $\cN=2$ SCFTs with weakly coupled supergravity duals:
\equ{\label{SFS3}
\log \cI_{S^2}(\omega) = \rmi\frac{F_{S^{3}}}{\pi } \frac{\varphi^{2}}{\omega}\,.
}
Similar universal relations between partition functions on different manifolds can be shown for the holographic duals of supersymmetric magnetic black holes in AdS$_4$ \cite{Azzurli:2017kxo}, as well as for theories in other dimensions \cite{Bobev:2017uzs,Benini:2015bwz}.

Note that until now we have only imposed the supersymmetric limit \eqref{eq:susylimit}. We emphasize that when this limit is imposed  the zeroes of the metric function $\Delta_r$ are generically not real and thus the Lorentzian solution has a naked singularity and causal pathologies. There is a special value of the parameters, however,  for which the supersymmetric solution is extremal, i.e., has vanishing temperature, and is a regular black hole  in Lorentzian signature. To obtain this supersymmetric and extremal black hole one must set
\equ{\label{constrJQ}
m=a(1+a)\sqrt{2+a}\qquad \Rightarrow  \qquad J=\frac{Q}{2}\(\sqrt{1+4G_{(4)}^{2}Q^{2}}-1\)\,,
}
leaving only one independent physical parameter, which we take to be the charge $Q$. The Bekenstein-Hawking entropy of the BPS black hole is real and given by 
\equ{\label{BHentropy}
S^{\rm BPS}_{\text{BH}}=\frac{\pi }{2G_{(4)}}\(\sqrt{1+4G_{(4)}^{2}Q^{2}}-1\)\,.
}
To account for the entropy  of the BPS black hole microscopically it is sufficient to establish the more general relation \eqref{SFS3} in field theory. As shown above, performing a Legendre transform and imposing the extremal relation \eqref{constrJQ} will automatically reproduce the entropy  \eqref{BHentropy}. In Section~\ref{sec:Microscopic entropy} we show how this can be done for a class of 3d SCFTs arising from M5-branes. 


\section{Uplifts and universality}
\label{sec:Uplifts and universality}

The label ``universal'' for the BPS black hole reviewed above refers to the fact that it can be embedded into string or M-theory in infinitely many ways. This is reflected in the choice of internal manifold and fluxes used in the uplift to 10d or 11d. One unifying feature of these uplifts is that the internal manifold has at least one $\U(1)$ isometry, dual to the $\U(1)_{R}$ R-symmetry of the 3d $\cN=2$ SCFT.  The precise information about the internal manifold and the fluxes on it determines the details of 3d SCFT dual. For any such realization of the black hole in string or M-theory equation \eqref{SFS3} holds and thus it provides a  universal relation between the superconformal index and the $S^{3}$ partition function of any 3d $\cN=2$ SCFT with a weakly coupled supergravity dual. Next, we discuss a number of explicit examples of such uplifts to string or M-theory.

\subsection{Wrapped M5-branes}
\label{sec:Wrapped M5-branes}

We begin by uplifting to M-theory on $M_{7}=S^{4}\times \Sigma_{3}$ where the $S^{4}$ is nontrivially fibered over a three-manifold $\Sigma_{3}$. To ensure the regularity of the solution $\Sigma_{3}$ has to be the three-dimensional hyperboloid with the constant curvature metric. One can also quotient the hyperboloid to produce a compact hyperbolic manifold. The uplift to eleven-dimensional supergravity can be performed using the results in  \cite{Donos:2010ax}. The metric reads
\eqss{
\label{upliftM5}
ds^2_{11} =2^{\frac{1}{6}}(1+\sin^2\nu)^{\frac{1}{3}}\Big(ds^2_4 +\tfrac{1}{\sqrt{2}}ds^2_{\Sigma_3}+\tfrac{1}{2}\Big[d\nu^2+&\frac{\sin^2\nu}{1+\sin^2\nu}\(d\psi-A\)^2\Big]\\
&\qquad+\frac{\cos^2\nu}{1+\sin^2\nu}\sum_{a=1}^{3}\(d\tilde{\mu}^a+\bar \omega^{ab}\tilde{\mu}^b\)^2\Big)\,,
}
where $ds^{2}_{4}$ is the black hole metric and $A$ is the gauge field in \eqref{BHsol}.\footnote{To match the conventions of  \cite{Donos:2010ax} to ours we need to set $g^2=\sqrt{2}$ and  rescale the gauge field as $gA_{(1)}^{\rm there} = A^{\rm here}$.} Here we have split the constrained coordinates $\mu^{i=1,\dots,5}$ on $S^{4}$  according to $\SO(3)\times \SO(2)\subset \SO(5)$, 
\begin{equation}
\begin{split}
\mu^a = \,&\cos(\nu) \, \tilde{\mu}^a\,, \qquad a=1,2,3\,,\\
 \mu^{\alpha}=\,&\sin(\nu) \, \tilde{\mu}^\alpha\,,\qquad \alpha=4,5\,,
\end{split}
\end{equation}
with $\sum_{i=1}^{5}(\mu^{i})^2=\sum_{a=1}^{3}(\tilde{\mu}^a)^2=\sum_{\alpha=4}^{5}(\tilde{\mu}^{\alpha})^2=1$ and use the explicit parametrization 
\begin{equation}
\begin{split}
\tilde{\mu}^1&=\cos\xi_1\,, \quad \tilde{\mu}^2=\sin\xi_1\cos\xi_2\,, \quad \tilde{\mu}^3=\sin\xi_1\sin\xi_2\,, \\
\tilde{\mu}^4&=\cos\psi\,, \qquad  \tilde{\mu}^5=\sin\psi\,,
\end{split}
\end{equation}
with coordinate ranges  $0\leq \nu\leq \pi/2$, $0\leq \psi<2\pi$, $0\leq \xi_{1}\leq \pi$, and $0\leq \xi_{2}<2\pi$. The metric on the hyperbolic manifold $ds^2_{\Sigma_3}$ is normalized such that $\hat{R}_{ab} = -g^2 \hat{g}_{ab}$. The four-form is given by
\eqss{\label{G4M5}
2^{\frac{3}{4}}G_4=\,&\frac{(5+\sin^{2}\nu)}{\(1+\sin^{2}\nu\)^{2}}\,  \epsilon_{abc}\epsilon_{\alpha\beta}\, D\mu^{b}\wedge D\mu^{c}\wedge\(\frac14\mu^{a} D\mu^{\alpha}\wedge D\mu^{\beta}+\frac16\mu^{\alpha}D\mu^{\beta}\wedge D\mu^{a}\)\\
\,&+\frac{\epsilon_{abc}}{(1+\sin^{2}\nu)} \Big[ \epsilon_{\alpha\beta} D\bar \omega^{ab}\wedge\( D\mu^{c}\wedge D\mu^{\alpha} \mu^{\beta}+\frac14  D\mu^{\alpha}\wedge D\mu^{\beta}\mu^{c}\)+2^{-\frac{3}{5}} F\wedge D\mu^{a}\wedge D\mu^{b}\mu^{c}\Big] \\
&\,-2^{-\frac14}\(\ast_{4}F\)\wedge \bar e^{a}\wedge D\mu^{a}+ 2^{-\frac{2}{5}}\ast_{7}\left[\(\ast_{4}F\)\wedge \bar e^{a}\right]\mu^{a}\,,
}
where we have defined
\equ{\label{Dmus}
D\mu^{a}=d\mu^{a}+\bar \omega^{ab} \mu^{b}\,,\qquad D\mu^{\alpha}=d\mu^{\alpha}+A \, \epsilon ^{\alpha \beta} \mu^{\beta}\,,
}
here  $\bar e^{a}$ and $\bar \omega^{ab}$ are, respectively,  the vielbein and spin connection for $ds^{2}_{\Sigma_3}$ and $F=dA$. If we restrict the four-form flux to the directions along the four-sphere we find
\equ{
2^{\frac34}G_4\Big|_{S^{4}}=-\, d\(\frac{\cos^{3}\nu}{1+\sin^{2}\nu}\)\wedge d\psi\wedge \text{dvol}(S^{2})\,,
}
with $\text{dvol}(S^{2})=\sin \xi_{2}\, d\xi_{1}\wedge d\xi_{2}$. This expression is useful for flux quantization which leads to the same result as for the uplift of the AdS$_4$ vacuum solution. This can be used to compute the free energy of the dual 3d $\mathcal{N}=2$ SCFT which gives  (see for example \cite{Gang:2014ema})
\equ{\label{FS3M5}
F_{S^{3}}= \frac{\pi }{2 G_{(4)}}=\frac{ \text{Vol}(\Sigma_{3})\, N^{3}}{3\pi}\,.
}
This eleven-dimensional solution corresponds to the backreaction of $N$ M5-branes wrapping the three-cycle $\Sigma_{3}\subset T^{*}\Sigma_{3}$ and spinning in the remaining $\Bbb R^{3}$. The wrapping on $\Sigma_{3}$ topologically twists the worldvolume theory on the worldvolume of the M5-branes, which amounts to turning on a background value for the  $\SO(3)\subset \SO(5)_{R}$ R-symmetry of the theory along $\Sigma_{3}$ to cancel its spin connection. This is manifested in the solution by the terms proportional to the spin connection $\bar \omega^{ab}$ in \eqref{Dmus} and in the first two terms in the second line of \eqref{G4M5}. Note that setting $A=0$ and replacing $ds^{2}_{4}$ by the AdS$_{4}$ metric reduces to the static wrapped M5-branes solution in \cite{Gauntlett:2000ng}, see also \cite{Gang:2014ema}.

The boundary 3d SCFT is then the theory obtained by twisted compactification of the 6d $(2,0)$ $A_{N-1}$ theory on $\Sigma_3$. This class of theories is often denoted $T_N[\Sigma_3]$ and referred to as theories of class $\cR$. Thus, we refer to the solution \eqref{upliftM5}-\eqref{G4M5} as a ``spinning black hole of class $\cR$.'' The on-shell action, or entropy function, for this black hole can be computed from \eqref{SFS3} and reads
\equ{\label{entropyfunctionR}
I_{E}({T_N[\Sigma_3]})\approx \frac{ N^{3}}{12 \rmi \pi^{2} \omega}\(\omega+2\pi \rmi\)^{2}\text{Vol}(\Sigma_{3})\,,
}
where we have imposed the constraint on $\varphi$ in the last equation in \eqref{eqsLeg}. Giving a microscopic account of the entropy of this black hole  amounts to reproducing \eqref{entropyfunctionR} from a field theory computation. We address this problem in Section~\ref{sec:Microscopic entropy}. 

\subsection{M2-branes}

A large class of three-dimensional $\mathcal{N}=2$ SCFTs arise on the worldvolume of M2-branes probing a conical CY four-fold. The prototypical example in this class is the ABJM theory which admits many generalizations in the form of $\mathcal{N}=2$ Chern-Simons matter theories. The holographic dual of these SCFTs is given by an AdS$_4\times \text{SE}_7$ Freund-Rubin solution of 11d supergravity, where SE$_{7}$ is the Sasaki-Einstein manifold that serves as the base of the conical CY four-fold. There are other generalizations of this construction in which there are internal fluxes on the internal manifold and the AdS$_4$ factor of the metric is warped, see for example \cite{Corrado:2001nv} and \cite{Gabella:2012rc}. In \cite{Gauntlett:2007ma} and \cite{Larios:2019lxq,Larios:2019kbw} it was shown that both of these classes of compactifications of M-theory admit a truncation to 4d $\mathcal{N}=2$ minimal gauged supergravity. This implies that the black hole of Section~\ref{sec:A universal spinning black hole in  AdS4} can be embedded in M-theory and interpreted as the backreaction of spinning M2-branes.

Here we show the explicit black hole solution realized as a deformation of the AdS$_4\times \text{SE}_7$ vacua of 11d supergravity. The 11d metric is
\begin{equation}
ds^2_{11} = \tfrac{1}{4} ds^2_4 + ds^2_{6}+ \left(d\psi+\sigma+\tfrac{1}{4}A\right)^2\,,
\end{equation}
and the 4-form flux is
\begin{equation}
G_4 = \tfrac{3}{8}\text{vol}_4 -\tfrac{1}{4} *_4 F \wedge J\,.
\end{equation}
Here $ds^2_4$ is the black hole metric in \eqref{BHsol}, $\text{vol}_4$ is its associated volume form, $A$ is the gauge field in \eqref{BHsol} and $F=dA$. The metric $ds^2_{6}$ is locally K\"ahler-Einstein and serves as the base for the SE$_7$ manifold with Reeb vector $\partial_{\psi}$. The K\"ahler 2-form on $ds^2_{6}$ is $J$ and $d\sigma = 2J$. The quantization condition of the four-form flux is the same as for the uplift of the AdS$_{4}$ vacuum. The free energy of the dual 3d $\mathcal{N}=2$ SCFT is given in terms of the volume of SE$_{7}$ by the well-known relation
\equ{
F_{S^{3}}=N^{3/2}\sqrt{\frac{2\pi^{6}}{27\text{Vol(SE$_{7}$)}}}\,,
}
where $N$ is the number of M2-branes. Using \eqref{SFS3} we arrive at the following expression for the large $N$ superconformal index of the same SCFT:
\equ{
\log \cI_{S^2} = \frac{\rmi}{\pi } \frac{\varphi^{2}}{\omega} N^{3/2}\sqrt{\frac{2\pi^{6}}{27\text{Vol(SE$_{7}$)}}}\,.
}
Deriving this expression with QFT methods would be very interesting and  lead to a microscopic account of the entropy of this family of black holes.

Perhaps the simplest example in this class of solutions is given by taking $\text{SE}_7=S^7/\Bbb Z_k$ in which case the boundary 3d SCFT is the ABJM theory \cite{Aharony:2008ug}. This theory has a global symmetry group with a Cartan subalgebra $\U(1)^4$. In the bulk supergravity this results in a more general class of black hole solutions with multiple electric charges. The entropy function for this class of black holes was identified in \cite{Choi:2018fdc} and a recent derivation from field theory was presented in \cite{Choi:2019zpz}. If one sets the fugacities $\Delta_{1,2,3,4}$ for the $\U(1)^{4}$ symmetry to be equal one recovers the universal black hole of Section~\ref{sec:A universal spinning black hole in  AdS4} and the results in \cite{Choi:2018fdc} and \cite{Choi:2019zpz} agree with \eqref{SFS3}.

\subsection{D2-branes}

Another way to engineer AdS$_4$ vacua with explicit 3d $\mathcal{N}=2$ field theory duals is to use D2-branes in massive IIA string theory, see \cite{Guarino:2015jca,Fluder:2015eoa}. As shown recently in \cite{Varela:2019vyd}, building on the results in \cite{Azzurli:2017kxo}, this class of solutions again admits a truncation to minimal gauged supergravity in four dimensions. Using these results we can find an explicit realization of the black hole in Section~\ref{sec:A universal spinning black hole in  AdS4} in massive IIA supergravity. The metric is
\begin{equation}
\begin{split}
ds^2_{10} = m^{\frac{1}{12}}2^{-\frac{5}{8}} (3+\cos2\alpha)^{\frac{1}{2}}(5+\cos2\alpha)^{\frac{1}{8}}\Big[ \tfrac{1}{3}ds^2_4+\tfrac{1}{2}d\alpha^2 +\frac{2\sin^2\alpha}{3+\cos2\alpha}ds^2_{\text{KE}_4}+\frac{3\sin^2\alpha}{5+\cos2\alpha}\hat{\eta}^2\Big]\,,
\end{split}
\end{equation}
where 
\begin{equation}
\hat{\eta} = d\psi + \sigma+ \tfrac{1}{3}A\,,
\end{equation}
and $d\sigma =2J $, where $J$ is the K\"ahler form on the four-dimensional space $ds^2_{\text{KE}_4}$ which admits a local K\"ahler-Einstein metric. The range of the angles $\alpha$ and $\psi$ is $0\leq\alpha\leq\pi$ and $0\leq\psi\leq2\pi$. The dilaton and  NS-NS 3-form are
\begin{equation}
\begin{split}
e^{\phi} &=  \frac{2^{\frac{1}{4}}}{m^{\frac{5}{6}}}\frac{(5+\cos2\alpha)^{\frac{3}{4}}}{3+\cos2\alpha}\,,\\
H_3 &= \frac{8}{m^{\frac{1}{3}}} \frac{\sin^{3}\alpha}{(3+\cos2\alpha)^2} J\wedge d\alpha+ \frac{1}{2\sqrt{3}m^{\frac{1}{3}}} \sin\alpha~ d\alpha \wedge *_{4} F\,.
\end{split}
\end{equation}
The RR fluxes are given by
\begin{eqnarray}
F_0 &=& m\,,\notag\\
m^{-\frac{2}{3}}F_2 &=& -\frac{4\sin^2\alpha \cos\alpha}{(3+\cos2\alpha)(5+\cos2\alpha)} J - \frac{3(3-\cos2\alpha)}{(5+\cos2\alpha)^2} \sin\alpha~ d\alpha\wedge \hat{\eta}  \notag\\
&& \qquad\qquad\qquad\qquad\qquad\qquad\qquad\qquad+\frac{\cos\alpha}{5+\cos2\alpha} F -\tfrac{1}{2\sqrt{3}}\cos\alpha *_4 F\,,\\
m^{-\frac{1}{3}}F_4 &=& \frac{2(7+3\cos2\alpha)}{(3+\cos2\alpha)^2} \sin^4\alpha~ \text{vol}_{KE_4} + \frac{3(9+\cos2\alpha)\sin^3\alpha\cos\alpha}{(3+\cos2\alpha)(5+\cos2\alpha)} J\wedge d\alpha\wedge \hat{\eta}+ \tfrac{1}{\sqrt{3}} \text{vol}_4\notag\,,\\
&&  -  \tfrac{1}{8}\sin2\alpha\Big(\frac{2\sin2\alpha}{3+\cos2\alpha} J + d\alpha\wedge \hat{\eta}\Big)\wedge F -  \tfrac{1}{4\sqrt{3}}\Big(\frac{4\sin^2\alpha}{3+\cos2\alpha} J + \frac{3\sin2\alpha}{5+\cos2\alpha}d\alpha\wedge \hat{\eta}\Big)\wedge *_4F\notag\,.
\end{eqnarray}
The solution is  interpreted as the backreaction of spinning D2-branes in massive IIA string theory. The boundary 3d SCFT is then the IR limit of the D2-brane world-volume theory. The simplest example in this class is the GJV theory \cite{Guarino:2015jca}. A generalization can be constructed by a certain ``descent'' procedure from 4d $\mathcal{N}=1$ quivers gauge theories \cite{Fluder:2015eoa}. This leads to the following relation between the free energy of the 3d theory and the conformal anomaly of the 4d SCFT 
\begin{equation}\label{eq:FS3D2}
F_{S^3} = \frac{2^{\frac{5}{3}}3^{\frac{1}{6}}\pi}{5} (nN)^{\frac{1}{3}}(a_{4d})^{\frac{2}{3}}\,,
\end{equation}
where $a_{4d}$ is the $a$-anomaly coefficient of the parent 4d theory and the relation is valid in the planar limit. Note that the free energy of these theories scales as $N^{5/3}$ which is characteristic for D2-branes in massive IIA string theory.  The map between field theory and supergravity quantities is provided by the quantization condition
\begin{equation}
N = \frac{1}{(2\pi\ell_s)^5} \frac{16}{3} \text{vol}(Y_5)\,, \qquad n = 2\pi\ell_s m\,.
\end{equation}
Here $Y_5$ is a five-manifold with a Sasaki-Einstein  metric
\begin{equation}
ds^2_{Y_5} = ds^2_{\text{KE}_4} +(d\psi+\sigma)^2\,,
\end{equation}
determining the AdS$_{5}$ IIB dual of the parent 4d theory. 

Using \eqref{SFS3} and \eqref{eq:FS3D2} we obtain a simple formula for the leading order in $N$ superconformal index for this large class of Chern-Simons matter theories:
\begin{equation}\label{cardyD2}
\log \cI_{S^2} = \rmi \frac{\varphi^{2}}{\omega}  \frac{2^{\frac{5}{3}}3^{\frac{1}{6}}}{5} (nN)^{\frac{1}{3}}(a_{4d})^{\frac{2}{3}}\,.
\end{equation}
Reproducing this index by field theory methods is an interesting open problem. 

\subsection{Wrapped D4-brane and $(p,q)$-fivebranes}
\label{sec:Wrapped D4- and p,q-fivebranes}

Another interesting class of 3d SCFTs are those obtained by twisted compactification of 5d SCFTs. Five-dimensional SCFTs  can be constructed in string theory from a system of D4-D8-O8 branes in massive type IIA string theory \cite{Intriligator:1997pq,Brandhuber:1999np}, studied holographically in \cite{Bergman:2012kr,Jafferis:2012iv}. Alternatively one can utilize $(p,q)$-fivebrane webs in type IIB string theory \cite{Aharony:1997ju,Kol:1997fv,Aharony:1997bh}, which was studied holographically in  \cite{DHoker:2016ysh,Uhlemann:2019ypp}. Upon a twisted compactification on a Riemann surface $\Sigma_{\fg}$ of genus $\fg$ one obtains a 3d $\cN=2$ theory whose $S^{3}$ partition function can be computed via the localization results of \cite{Crichigno:2018adf} where a universal relation to the free energy on $S^{5}$ was derived; we comment on this further in Section~\ref{sec:Outlook}. These 3d $\cN=2$ SCFTs admit a bulk description in terms of warped AdS$_4$ vacua of massive IIA or IIB supergravity, see for example \cite{Bobev:2017uzs} and \cite{Bah:2018lyv}. We conjecture that these supergravity compactifications should admit a consistent truncation to minimal 4d gauged supergravity. More specifically, we expect a truncation of massive IIA on $M_{6}=S^{4}\times_{w} \Sigma_{\fg>1}$ where the $S^{4}$ is fibered over the Riemann surface and a truncation of IIB supergravity on $M_{6}=S^{2}\times_{w} \Sigma \times_{w} \Sigma_{\fg}$, where $\Sigma$ is a Riemann surface with disc topology and the $S^{2}$ is fibered over $\Sigma_{\fg}$. Unfortunately these truncations have not been established in the literature. Nevertheless, our results suggest that they should exist and thus the universal black hole in Section~\ref{sec:A universal spinning black hole in  AdS4} can be embedded in these AdS$_4$ vacua. Establishing the universal formula \eqref{SFS3} for these 3d SCFTs would then account for the entropy of the corresponding black holes. This is certainly an interesting and nontrivial problem beyond the scope of this work.

\section{Microscopic entropy and class $\mathcal{R}$}
\label{sec:Microscopic entropy}

In this section, we study the universal spinning black hole in AdS$_{4}$ from a holographic perspective. We focus on the uplift to M-theory, discussed in Section~\ref{sec:Wrapped M5-branes},  on a manifold $M_{7}$ which is a fibration of a squashed $S^{4}$ over $\Sigma_{3}$. In this case the entropy of the black hole should be accounted for by the superconformal index of  3d $\cN=2$ theories of class $\cR$. We exploit the 3d-3d correspondence which relates the index to the partition function of complex Chern-Simons theory on $\Sigma_{3}$ and show that this is indeed the case.  Similar results for the uplift of supersymmetric magnetic black holes without angular momenta were obtained in \cite{Gang:2018hjd, Gang:2019uay,Bae:2019poj}. 

\subsection{The 3d-3d correspondence}

Consider the 6d $(2,0)$  $A_{N-1}$ theory arising on the worldvolume of $N$ M5-branes. The theory can be placed on $\Bbb R^3\times M_3$ with $M_3$ a generic three-manifold while preserving four supercharges. This is achieved by performing a topological twist  using the $\SO(3)\subset \SO(5)_R$ R-symmetry to cancel the spin connection on $M_3$. Taking $M_{3}$ to be compact at low enegies one obtains a 3d $\cN=2$ SCFT of class $\cR$ denoted by $T_N[M_3]$ \cite{Dimofte:2011ju,Cecotti:2011iy,Dimofte:2011py}. The 3d-3d correspondence \cite{Dimofte:2011ju,Cecotti:2011iy,Dimofte:2011py,Dimofte:2010tz,Terashima:2011qi,Dimofte:2011jd} maps the supersymmetric partition function of  $T_{N}[M_{3}]$ on a curved background $B$ to topological invariants of  $M_3$, see \cite{Dimofte:2016pua} for a review and further references.  For $B=S^{3}_{b}/\Bbb Z_{k\geq 1}$ and for $B=S^{2}_{\omega}\times S^{1}$ the corresponding topological invariant is given by a complex Chern-Simons (CS) partition function on $M_3$.\footnote{Topologically twisted indices in the 3d-3d correspondence context were studied in \cite{Gukov:2017kmk,Gukov:2016gkn,Lee:2013ida}.} More precisely we have the following relation  between partition functions
\equ{\label{3d3dcorr}
Z_{B}(T_{N}[M_{3}])\stackrel{\text{3d-3d}}{=}Z_{N}^{CS}(\hbar, \tilde \hbar;M_{3})\,,
}
where the path integral for the Chern-Simons theory is defined as
\eqss{\label{CSPI}
Z_{N}^{CS}(\hbar, \tilde \hbar;M_{3})=\int \cD\cA\cD\bar \cA\, e^{\frac{\rmi }{2\hbar} S_{CS}[\cA;M_{3}] + \frac{\rmi }{2\tilde \hbar}S_{CS}[\bar \cA;M_{3}]}\,,
}
where $S_{CS}[\cA;M_{3}]=\int_{M_{3}}\text{Tr}\,\(\cA \, d \cA+\frac23 \cA^{3}\)$ is the usual Chern-Simons action, with $\cA,\bar \cA$ complex connections, valued in the Lie algebra of the group $\SL(N,\Bbb C)$ and $\hbar,\tilde \hbar$ are two complex parameters, usually written as
\equ{
\frac{4\pi}{\hbar}=k+\rmi s\,,\qquad \frac{4\pi}{\tilde \hbar}=k-\rmi s\,.
}
It follows from standard arguments of real CS theory that $k\in \Bbb Z$  while $s$ is constrained only by unitarity to be either real or purely imaginary  \cite{Witten:1989ip}. The choice of $B$ in the LHS of \eqref{3d3dcorr} is encoded in the RHS in the values of  $\hbar,\tilde \hbar$. For the superconformal index of interest here we have $B=S^{2}_{\omega}\times S^{1}$. This corresponds to setting $\hbar=\rmi\omega$ and $\tilde \hbar=-\rmi \omega$ (i.e. $k=0$ and $s=-\frac{4\pi}{\omega}$).  From now on we set $k=0$ so we have
\equ{\label{PICS}
Z_{S^{2}_{\omega}\times S^{1}}(T_{N}[M_{3}])=Z_{N}^{CS}(\rmi \omega, -\rmi \omega;M_{3}) =  \int \cD\cA \cD\bar \cA\, e^{\frac{\rmi}{\omega}\text{Im}\, S_{CS}[\cA;M_{3}]}\,.
}
Note the parameter $\omega$ in the CS theory has been identified with the rotational chemical potential in the superconformal index, which by the AdS/CFT correspondence is also identified with the potential in the supergravity theory as defined in \eqref{eq:omphidefsugra}. 

To compare the field theory calculation to the result from supergravity \eqref{entropyfunctionR} we must evaluate the CS partition function at large $N$. Without any other simplifying assumptions this is a nontrivial task. Note, however, that if $\omega$ is analytically continued to imaginary values one can define a semi-classical limit, $|\omega|\to 0$, of the CS theory which translates into a Cardy-like limit of the superconformal index of $T_{N}[M_{3}]$. In this regime the path integral \eqref{PICS}  can be evaluated at the perturbative level by expanding around  saddles of the CS action, which are given by flat connections, $\cA^{(\alpha)}$, with the dominant contribution determined by the value of $\text{Im} \, S_{CS}[\cA^{(\alpha)};M_{3}]$.

Guided by the supergravity solution in Section~\ref{sec:Wrapped M5-branes} we are interested in the case where $M_{3}$ is a smooth quotient of the three-dimensional hyperboloid, $\Sigma_{3}=\mathbb{H}^3/\Gamma$. In this case the vielbein, $e$, and the spin connection, $w$, of $\Sigma_{3}$ can be rewritten as flat connections for $\SL(2,\Bbb C)$ by taking the complex combinations $w\pm \rmi e$.  One can then use these two geometric connections to construct flat connections of $\SL(N,\Bbb C)$, explicitly given by
\equ{A^{\text{(geom)}}_{N}=\rho_{N}\cdot(w+\rmi e)\,,\qquad A^{\overline{\text{(geom)}}}_{N}=\rho_{N}\cdot(w-\rmi e)\,,}
where $\rho_{N}$ is the $N$-dimensional irreducible representation of $\SL(2,\Bbb C)$. An important property of these geometric flat connections is that
\equ{
\text{Im}\(S_{CS}[\cA^{\overline{\text{(geom)}}}_{N};\Sigma_{3}]\)\leq \text{Im}\(S_{CS}[\cA^{(\alpha)}_{N};\Sigma_{3}]\) \leq \text{Im}\(S_{CS}[\cA^{\text{(geom)}}_{N};\Sigma_{3}]\)\,,
}
where the lower and upper inequalities are saturated only for $\cA^{(\alpha)}_{N}=\cA^{\overline{\text{(geom)}}}_{N}$ and $\cA^{{\text{(geom)}}}_{N}$, respectively, and thus dominate the saddle-point approximation.\footnote{A subtle point is that to fully capture the superconformal theory $T_{N}[M_{3}]$ one should also include reducible flat connections \cite{Chung:2014qpa}. Here, as in \cite{Gang:2018hjd,Gang:2019uay,Bae:2019poj}, we assume that for the purposes of holography, to leading order in $N$, it is sufficient to consider only irreducible connections. Our results below give further evidence for this assumption. } Let us approach the origin from below, $\omega\to \rmi0^{-}$. Then,  the flat connection $\cA^{\overline{\text{(geom)}}}_{N}$ is the dominant saddle and by a standard semi-classical approximation the partition functions takes the form\footnote{Approaching the origin from above, $\omega \to \rmi 0^{+}$, the dominant saddle is instead the flat connection $\cA^{\text{(geom)}}_{N}$. The final answer is the same and the limit is well defined. } 
\equ{\label{pertexp}
Z_{N}^{CS}(\rmi \omega, -\rmi \omega;\Sigma_{3}) \xrightarrow{\omega \to \rmi 0^{-}}\exp\left\{\frac{2\rmi}{\omega}\, \text{Im}\(S_{0}^{\overline{\text{(geom)}}}+\omega S_{1}^{\overline{\text{(geom)}}}+\omega^{2}S_{2}^{\overline{\text{(geom)}}}+\dots\)\right\}\,,
}
where the first coefficient  is given by the value of the classical action, $S_{0}^{\overline{\text{(geom)}}}=\frac12 S_{CS}[\cA^{\overline{\text{(geom)}}}_{N};\Sigma_3]$, the coefficient $S_{1}^{\overline{\text{(geom)}}}$ is a 1-loop correction, etc.\footnote{Here we have included a $\log \omega$ term in the ellipses in \eqref{pertexp}. This term  will not be important in the large $N$ limit.}   In principle, these coefficients can  be computed via higher loop Feynman diagrams, although this may not be very efficient.  To compare with supergravity we must evaluate \eqref{pertexp} in the large $N$ limit.

\subsection{Large $N$}

As summarized in \cite{Gang:2014ema} the classical and one-loop contributions in  \eqref{pertexp}, $S_{0,1}^{\overline{\text{(geom)}}}$, and their conjugates, can be computed directly. The two-loop coefficient  is more involved. In the large $N$ limit, however, its value can be conjectured with input from holography by comparing the  on-shell action for the AdS$_{4}(b)$ solution to the free energy of the 3d SCFT on the squashed $S^{3}_{b}$. This leads to the following results \cite{Gang:2014ema}:
\eqss{\label{Sbargeom}
\text{Im}[S_{0}^{\overline{\text{(geom)}}}]=\,&- \text{Im}[S_{0}^{{\text{(geom)}}}]\approx-\frac{1}{6} \text{Vol}(\Sigma_{3}) \,N^{3 }\,,\\
\text{Re}[S_{1}^{\overline{\text{(geom)}}}]=\,&\text{Re}[S_{1}^{{\text{(geom)}}}]\approx -\frac{1}{6\pi} \text{Vol}(\Sigma_{3}) \,N^{3}\,,\\
\text{Im}[S_{2}^{\overline{\text{(geom)}}}]=\,&-\text{Im}[S_{2}^{{\text{(geom)}}}]\approx\frac{1}{24\pi^{2}} \text{Vol}(\Sigma_{3})\, N^{3}\,,
}
to leading order in $N$ and  all $S^{(\alpha)}_{n\geq3}$ are subleading. Note that the result for $S_{2}^{\overline{\text{(geom)}}}$, as well as the vanishing of $S^{(\alpha)}_{n\geq3}$, in the large $N$ limit are conjectured results. Nevertheless, there is convincing evidence that these conjectures are true \cite{Gang:2014ema}. Plugging the results in \eqref{Sbargeom} into  \eqref{pertexp}  and using  the 3d-3d correspondence relation in \eqref{PICS}, we find the following expression for the superconformal index of the theory $T_N[M_3]$ in the large $N$ limit 
\equ{\label{indexQFT}
\log \, Z_{S^2_\omega\times S^1}(T_N[\Sigma_3])\approx - \frac{ N^{3}}{12 \rmi \pi^{2} \omega}\(\omega+2\pi \rmi\)^{2}\text{Vol}(\Sigma_{3})\,.
}
This is the main result of our analysis for theories of class $\cR$. Due to the holographic dictionary \eqref{eq:IZI} this  should match (minus) the on-shell action of the spinning black hole. Indeed, comparing to   \eqref{entropyfunctionR} we find precise agreement.  As discussed around (\ref{eq:Iintro})-(\ref{eq:SBHintro}), to account for the entropy of the black hole one has to analytically continue $\omega$ to the complex plane and Legendre transform to an ensemble with fixed  charge $Q$ and angular momentum $J=J(Q)$. Implementing this procedure for the superconformal index in \eqref{indexQFT} automatically reproduces the Bekenstein-Hawking entropy in \eqref{BHentropy} with the appropriate quantized value of Newton's constant  from \eqref{FS3M5}.

There is an important subtlety in the calculation above. In the supergravity construction it was assumed that the hyperbolic manifold $\Sigma_3$ on which the M5-branes are wrapped is compact and admits a smooth metric. On the other hand the large $N$ results for theories of class $\cR$ in \cite{Gang:2014ema} are derived for three-manifolds which are knot complements and thus have some defects. Given that we find a nontrivial agreement between the supergravity and the large $N$ field theory calculations it is natural to conjecture that the contribution from the singularities of the metric on $\Sigma_3$ will be subleading in the $1/N$ expansion. This was also assumed in similar holographic calculations in \cite{Gang:2018hjd,Gang:2014ema,Gang:2019uay,Bae:2019poj}. It is certainly desirable to understand this issue better.

We note that  \eqref{indexQFT} was derived in the  Cardy-like limit, $|\omega| \to 0$,  and large $N$ while \eqref{entropyfunctionR} is valid more generally for any $\omega$ and large $N$. It is important to keep in mind that in a given charge sector there may be more BPS states than the  black hole discussed above. These other states, such as multi-center black holes if they exist, would also contribute to the index. The match shown above indicates that at least at the perturbative level in $|\omega|$, and in the large $N$ limit,  the (single-center)  black hole solution dominates and thus the index correctly captures its entropy. It would be interesting to study corrections beyond the Cardy-like and large $N$ limits and the corresponding supergravity interpretation of such corrections.

\section{Outlook}
\label{sec:Outlook}

The most pressing question stemming from our work is to  establish the 3d Cardy-like formula \eqref{IS3intro} by pure field theory methods. This may be possible, for instance, for 3d $\cN=2$ Chern-Simons quiver gauge theories along the lines of a similar universal relation discussed in  \cite{Azzurli:2017kxo}  for the topologically twisted index. As   discussed above, one should bear in mind that \eqref{IS3intro} holds provided the  black hole is the dominant contribution to the index and we have given an example where this holds for large $N$. It would therefore be very interesting to study corrections to the superconformal index beyond the Cardy-like and large $N$ limits. Corrections in the $1/N$ expansion  have been studied for twisted partition functions of class $\cR$ in \cite{Gang:2019uay}. We should also emphasize that in Section~\ref{sec:Microscopic entropy} we have used a number of results on the large $N$ limit of the 3d-3d correspondence discussed in \cite{Gang:2014ema,Gang:2018hjd}. Some of these results are rigorously derived but others are still a conjecture. Understanding the results in \cite{Gang:2014ema,Gang:2018hjd, Gang:2019uay,Bae:2019poj} more rigorously is certainly  interesting, especially in the context of holography.

An interesting class of 3d $\cN=2$ theories, which is much less explored, is obtained by twisted compactification of 5d SCFTs on a Riemann surface $\Sigma_{\fg}$, and whose various partition functions are accessible via supersymmetric localization \cite{Crichigno:2018adf,Hosseini:2018uzp}. In particular, the $S^{3}$  partition function was  computed  in \cite{Crichigno:2018adf}, and one can show that for $\fg>1$ and large $N$ there is a universal relation, $F_{S^{3}}=-\frac89(\fg-1)F_{S^{5}}$, with $F_{S^5}$ the free energy on the round five-sphere, as predicted by supergravity \cite{Bobev:2017uzs}. Combining this  with \eqref{SFS3} leads to the following prediction for the partition function of the 5d SCFT on $S^1\times S^2_{\omega}\times \Sigma_{\fg>1}$ in large $N$ limit:
\equ{
\log Z_{S^1\times S^2_{\omega}\times \Sigma_{\fg>1}}\approx  \frac89 (\fg-1)\frac{F_{S^5}}{ \rmi \pi } \frac{\varphi^{2}}{\omega}\,.
}
It would be interesting to establish this relation in field theory by directly computing the partition function of 5d $\cN=1$ gauge theories on $S^{1}\times S^{2}_{\omega}\times \Sigma_{\fg>1}$, with a partial topological twist on $\Sigma_{\fg>1}$. 

As emphasized in \cite{Bobev:2017uzs} the universality argument holds for any solution to minimal gauged supergravity in arbitrary dimension. In particular, uplifting the supersymmetric spinning black hole in AdS$_6$ of \cite{Chow:2008ip} to 10d or 11d predicts interesting Cardy-like formulas for the corresponding 5d SCFTs at large $N$. The on-shell action for this black hole was recently evaluated in \cite{Cassani:2019mms}, showing that it reproduces the entropy function introduced in \cite{Choi:2018fdc}.  Universality then predicts that, in the regime in which the black hole is the dominant contribution, the index of 5d SCFTs with a weakly coupled gravity dual is given by\footnote{Here we have used the holographic relation $F_{S^{5}}=-\frac{\pi^{2}}{3G_{(6)}}$ for AdS$_{6}$ vacua.} 
\equ{\label{5dcardy}
\log Z_{S^{1}\times S^{4}_{\omega_{1},\omega_{2}}}\approx-\frac{\rmi}{\pi}F_{S^{5}}\frac{\varphi^{3}}{\omega_{1}\omega_{2}}\,,
}
where $\varphi$ and $\omega_{1,2}$ are fugacities for the R-symmetry and the two rotations of $S^{4}$, respectively,  subject to the constraint $\omega_{1}+\omega_{2}-3\varphi=\pm 2\pi \rmi$. In the case of 5d SCFTs arising from D4-D8-O8 branes in massive type IIA string theory this formula was established by localization methods in \cite{Choi:2019miv} in the Cardy-like limit $|\omega_{1,2}|\ll0$. We claim here that \eqref{5dcardy} holds for {\it any} SCFT with a weakly coupled gravity dual,  for instance those arising from $(p,q)$-fivebranes in IIB string theory. It  would be interesting to establish this by pure field theory methods and determine the precise regime of dominance of the universal black hole solution.

Finally, supersymmetric asymptotically AdS$_4$ black holes with angular momentum and both electric and magnetic charges have been recently constructed in \cite{Hristov:2018spe,Hristov:2019mqp}. It would be interesting to study the corresponding universal behavior of the entropy function for these solutions, whose entropy should be captured by the 3d topologically twisted index, refined by angular momentum. 

\bigskip
\bigskip

\noindent \textbf{Acknowledgements }
\bigskip

We would like to thank Davide Cassani, Anthony Charles, Fir\dh rik Freyr Gautason, Seyed Morteza Hosseini,  Kiril Hristov, Vincent Min, and Brian Willett for useful discussions. The work of NB is supported in part by an Odysseus grant G0F9516N from the FWO and the KU Leuven C1 grant ZKD1118 C16/16/005. PMC is supported by Nederlandse Organisatie voor Wetenschappelijk Onderzoek (NWO) via a Vidi grant and is also part of the Delta ITP consortium, a program of the NWO that is funded by the Dutch Ministry of Education, Culture and Science (OCW). PMC would like to thank KU Leuven for hospitality during part of the project. Both of us would like to thank the Mainz
Institute for Theoretical Physics (MITP) of the Cluster of Excellence PRISMA+ (Project ID 39083149) for hospitality during the initial stages of this project.

\appendix

\section{Boundary Killing spinors}
\label{sec:Boundary Killing spinors}

Three-dimensional backgrounds preserving supersymmetry can be constructed by coupling the field theory to new-minimal supergravity. The bosonic content consists of the vielbein $e_\mu^a$, a gauge field $A^{\text{nm}}_{\mu}$, a conserved vector field $V^{\text{nm}}_{\mu}$, and a scalar $H$. We denote the two complex supersymmetry generators of 3d $\cN=2$ supersymmetry by $\zeta,\tilde \zeta$. Each is a doublet of the $\SU(2)$ rotation group of $S^2$ and carry $\U(1)_R$ R-charges $(1,-1)$, respectively.  The Killing spinor equations read \cite{Closset:2012ru}
\eqss{
(\nabla_{\mu}- \rmi A^{\text{nm}}_{\mu})\zeta =\,&- \frac12 H\gamma_{\mu} \zeta- \rmi V_{\mu}^{\text{nm}}\zeta-\frac12\epsilon_{\mu\nu\rho}V^{\nu}_{\text{nm}}\gamma^{\rho}\zeta\,, \\
(\nabla_{\mu}+\rmi A^{\text{nm}}_{\mu})\tilde \zeta =\,&- \frac12 H\gamma_{\mu} \tilde \zeta+\rmi V_{\mu}^{\text{nm}}\tilde \zeta+\frac12\epsilon_{\mu\nu\rho}V^{\nu}_{\text{nm}}\gamma^{\rho}\tilde \zeta \,,
}
where $\nabla_{\mu}\zeta=\partial_{\mu} \zeta+\frac14 w_{\mu}^{ab}\gamma_{ab}\zeta$, with $w_{\mu}^{ab}$ the spin connection and $\gamma_{a}$ are the Dirac gamma matrices with flat indices. We work in the representation $\gamma_{a}=(\sigma_1,\sigma_2,\sigma_3)$ with $\sigma_a$ the Pauli matrices. In the conformal case  $H$ is set to zero and $A_{\mu}^{\text{nm}}-\frac12 V_{\mu}^{\text{nm}}$ is pure gauge. The combination $A_{\mu}^{\text{cs}}\equiv A_{\mu}^{\text{nm}}-\frac32 V_{\mu}^{\text{nm}}$ remains and is identified with the R-symmetry gauge field of conformal supergravity, which in holography is fixed by the boundary value of the bulk supergravity solution \cite{Klare:2012gn}. We are interested in solutions to these equations for the asymptotic metric and gauge field  \eqref{bdrymetriclocal}. Thus we set  $A_{\mu}^{\text{cs}}=-\rmi \alpha \delta_{\mu}^{3}$ and a consistent choice is 
\equ{
A_{\mu}^{\text{nm}} =A_{\mu}^{\text{cs}}-\frac32 \rmi\delta_{\mu}^{3}\,, \qquad V_{\mu}^{\text{nm}} = - \rmi \,\delta_{\mu}^{3}\,.
} 
The Killing spinor equations can then be written as
\eqs{
(\nabla_\mu- \rmi A_{\mu}^{\text{cs}})\zeta = \frac{1}{2}\gamma_\mu \gamma_3 \zeta\,,\qquad 
(\nabla_\mu+ \rmi A_{\mu}^{\text{cs}})\tilde \zeta = -\frac{1}{2}\gamma_\mu \gamma_3 \tilde \zeta\,,
}
with solutions
\begin{equation}
\begin{aligned}[c]
\zeta^{(1)}=\,&e^{ \tau(\alpha+\frac{1}{2})}e^{\frac{\rmi \hat\phi}{2}}\(\begin{matrix}\cos \frac{\hat \theta}{2}\\ \sin \frac{\hat \theta}{2}\end{matrix}\)\,,\\
\tilde \zeta^{(1)}=\,&e^{- \tau(\alpha+\frac{1}{2})}e^{\frac{\rmi \hat\phi}{2}}\(\begin{matrix}\cos \frac{\hat \theta}{2}\\- \sin \frac{\hat \theta}{2}\end{matrix}\)\,,
\end{aligned}
\qquad
\begin{aligned}[c]
& \zeta^{(2)}=e^{ \tau(\alpha+\frac{1}{2})}e^{-\frac{\rmi \hat\phi}{2}}\(\begin{matrix}\sin \frac{\hat \theta}{2}\\ - \cos \frac{\hat \theta}{2}\end{matrix}\)\,,\\
& \tilde \zeta^{(2)}=e^{ -\tau(\alpha+\frac{1}{2})}e^{-\frac{\rmi \hat\phi}{2}}\(\begin{matrix}\sin \frac{\hat \theta}{2}\\  \cos \frac{\hat \theta}{2}\end{matrix}\)\,.
\end{aligned}
\end{equation}

We  see  from \eqref{idphihat}  that demanding regularity of the bulk solution at the horizon implies the following transformations of the  spinors at the boundary:
\begin{equation}
\begin{aligned}[c]
\zeta^{(1)}\to\,&  e^{\frac{\beta}{2}( 1 +\Omega +2\alpha)}\zeta^{(1)}\,,\\
\tilde \zeta^{(1)}\to\,& e^{\frac{\beta}{2}(- 1 +\Omega -2\alpha)}\tilde \zeta^{(1)}\,,
\end{aligned}
\qquad
\begin{aligned} 
\zeta^{(2)}\to\,& e^{\frac{\beta}{2}( 1 - \Omega +2\alpha)}\zeta^{(2)}\,,\\
\tilde  \zeta^{(2)}\to\,& e^{\frac{\beta}{2}(- 1 -\Omega -2\alpha)}\tilde \zeta^{(2)}\,.
\end{aligned}
\end{equation}
With a suitable choice of the background R-symmetry gauge field it is  possible to preserve two supersymmetries of opposite R-charge. For instance, choosing $\alpha$ so that 
\equ{\label{constOmega_alpha}
\beta(1+\Omega+2 \alpha)=2\pi \rmi \, n\,,\qquad n \in \Bbb Z\,,
}
two of the spinors become periodic or anti-periodic, depending on whether $n$ is even or odd:
\equ{
\zeta^{(1)}\to  e^{\rmi\pi n}\, \zeta^{(1)}\,,\qquad \tilde \zeta^{(2)}\to  e^{-\rmi\pi n}\, \tilde \zeta^{(2)}\,.
}
The remaining Killing spinors are generically neither periodic nor anti-periodic and the corresponding supersymmetries are broken.  At this point we make contact with the black hole solution by setting $\alpha=-\Phi$, with $\Phi$ the chemical potential in the bulk supergravity solution. Then, as a consequence of the bulk relation \eqref{constOmegaPhi}, the constraint \eqref{constOmega_alpha} is satisfied with  $n=\pm1$ and the two preserved Killing spinors are anti-periodic. For $n=0$ instead the spinors are periodic and the bulk solution is the AdS$_4$ vacuum.

\section{Superconformal index}
\label{sec:SCindex}

Consider a 3d $\cN=2$ SCFT in Euclidean signature, radially quantized on $S^2\times \Bbb R$.  We denote the Poincar\'e supercharges by $\cQ_{1},\cQ_2,\tilde \cQ_1,\tilde \cQ_2$.  Their conjugates,  $\cQ_{1}^\dagger,\cQ_2^\dagger,\tilde \cQ_1^\dagger,\tilde \cQ_2^\dagger$, are identified with the superconformal charges.  The global charges of these supersymmetries are shown in Table~\ref{TableQs}, where $\Delta$ is the dilation operator, $j_{3}$ is the Cartan of $\SU(2)$ and $R$ is the generator of $\U(1)_{R}$. 
\begin{table}[htp]
\begin{center}
\begin{tabular}{c|c|c|c|c||c|c|c|c}
&$\cQ_1 $ & $\cQ_2$ & $\tilde \cQ_1$ & $\tilde \cQ_2$ & $\cQ_1^\dagger $ & $\cQ_2^\dagger$ & $\tilde \cQ_1^\dagger$ & $\tilde \cQ_2^\dagger$ \\ \hline 
$\Delta $ & $ \frac12$ & $\frac12$ & $\frac12$  & $\frac12$ & $-\frac12$ & $-\frac12$ & $-\frac12$ & $-\frac12$  \\
$R $ & $ 1$ & $1$ & $-1$ & $-1$ & $ -1$ & $-1$ & $1$ & $1$  \\
$j_3$ & $ -\frac12$ & $ \frac12$ & $ -\frac12$  & $ \frac12$ & $ \frac12$ & $ -\frac12$  & $ \frac12$ & $ -\frac12$    
\end{tabular}
\end{center}
\caption{Global charges of supersymmetry generators of the 3d $\cN=2$ superconformal algebra.}
\label{TableQs}
\end{table}%

One now chooses a supercharge, say $\cQ_1$, and its conjugate $\cQ_1^\dagger$. It follows from the superconformal algebra that
\equ{
\{\cQ_1^\dagger,\cQ_1\}= \Delta-R-j_3\,.
}
Note the combination $\Delta+j_3$ commutes with $\cQ_1,\cQ_1^\dagger$. Then, one defines the index
\equ{
\cI_{S^2}(\omega)=\text{Tr}\,  e^{\rmi \pi R}\, e^{\frac12 \gamma \{\cQ_1^\dagger,\cQ_1\}}e^{\frac 12\omega(\Delta+j_3)}\,,
}
where the trace is evaluated over the Hilbert space of the theory quantized on $S^{2}$. The $  e^{\rmi \pi R}$ factor anticommutes with $\cQ_1$ and $\cQ_1^\dagger$ and thus acts as the more standard $(-1)^F$, making this quantity an index which receives only contributions from states annihilated by $\cQ_1$ and $\cQ_1^\dagger$. Then, only states with $\Delta=R+j_3$ contribute.  In particular, the index is independent of the parameter $\gamma$. Of course, one may choose another supercharge and define the corresponding index but this will contain equivalent information. Using the anti-commutation relation above we can write
\equ{
\cI_{S^2}(\omega)=\text{Tr}\,  e^{ R (\rmi \pi -\frac{\gamma}{2})}\, e^{\frac12 \Delta (\gamma+\omega)}e^{\frac12 j_3 (\omega-\gamma)}\,.
}
We may now use the freedom to choose $\gamma$ to set $\gamma=-\omega$, after which 
\equ{\label{defI}
\cI_{S^2}(\omega)=\text{Tr}\,  e^{ \varphi R}\,e^{\omega j_3 }=\sum_{Q,J}\Omega(Q,J)\,e^{\varphi Q} e^{\omega J}\,, \qquad  \varphi \equiv \rmi \pi +\frac12\omega\,,
}
where $\Omega(Q,J)$ is the degeneracy of states with R-charge $Q$ and angular momentum $J$.  This can be seen as a path integral, $Z_{S^{2}_{\omega}\times S^{1}}$, over the background described above.

Note that defining the shifted fugacity $\hat \omega=\omega-2\pi \rmi$ we can also write the index in the form
\equ{
\cI_{S^2}(\hat \omega)=\text{Tr}\,  (-1)^{F}\, e^{ \frac12 \hat \omega R}\,e^{\hat \omega j_3 }=\text{Tr}\,  (-1)^{F}\, e^{ \frac12 \hat \omega (\Delta+j_{3})}\,,
}
where we used $e^{-2\pi \rmi \, j_{3}}= (-1)^{F}$, as a consequence of spin-statistics. This is the more standard form of the index (see, e.g., \cite{Imamura:2011su}). For our purposes it is more convenient to work with the fugacity $\omega$ and the form \eqref{defI}.

\bibliography{BH_class_R} 
\bibliographystyle{JHEP}

\end{document}